# Edge-based 2D α-In$_2$Se$_3$-MoS$_2$ ferroelectric field effect device

Debopriya Dutta[1], Subhrajit Mukherjee[1], Michael Uzhansky[1], Pranab K. Mohapatra[2], Ariel Ismach[2] and Elad Koren[1*]

[1]*Nanoscale Electronic Materials and Devices Laboratory, Faculty of Materials Science and Engineering, Technion - Israel Institute of Technology, Haifa, 3200003, Israel.*

[2]*Department of Materials Science and Engineering, Tel Aviv University, Ramat Aviv, Tel Aviv, 6997801, Israel.*

[*]Email – eladk@technion.ac.il

## Abstract -

Heterostructures based on two dimensional (2D) materials offer the possibility to achieve synergistic functionalities which otherwise remain secluded by their individual counterparts. Herein, ferroelectric polarization switching in α-In$_2$Se$_3$ has been utilized to engineer multilevel non-volatile conduction states in partially overlapping α-In$_2$Se$_3$-MoS$_2$ based ferroelectric semiconducting field effect device. In particular, we demonstrate how the intercoupled ferroelectric nature of α-In$_2$Se$_3$ allows to non-volatilely switch between *n-i* and *n-i-n* type junction configurations based on a novel edge state actuation mechanism, paving the way for sub-nanometric scale non-volatile device miniaturization. Furthermore, the induced asymmetric polarization enables enhanced photogenerated carriers' separation, resulting in extremely high photoresponse of ~ 1275 A/W in the visible range and strong non-volatile modulation of the bright A- and B- excitonic emission channels in the overlaying MoS$_2$ monolayer. Our results show significant potential to harness the switchable polarization in partially overlapping α-In$_2$Se$_3$-MoS$_2$ based FeFETs to engineer multimodal, non-volatile nanoscale electronic and optoelectronic devices.

## Introduction –

Transition metal dichalcogenides (TMDCs) in general and MoS$_2$ in particular have been among the most sought after two dimensional (2D) semiconductor materials due to their excellent ambient stability[1,2], high mobility[3,4], high on-off ratio[5,6] and superior photo-responsivity[7,8]. Moreover, their considerable band gaps place them as promising candidates for fabricating multimodal optoelectronic devices[9]. Among the ways to non-volatilely control the conduction states in operational field effect devices based on 2D materials, fabricating heterostructures with ferroelectric materials is a promising direction. Such unique device geometry opens the realm for various advanced applications in the form of field effect devices (FETs)[10], transducers[11], sensors[12], non-volatile memory[13], neuromorphic[14,15], energy harvesting devices[16] etc. While traditional piezoelectric materials based on BaTiO$_3$, BiFeO$_3$, and Pb(Zr, Ti)O$_3$ have been used in such applications[17–20], the presence of truncated interfaces which are laden with interfacial electronic states strongly degrades the electronic characteristics compared to the pristine material's form, particularly when scaling down the device dimensions. To overcome such challenges, the requirement for a ferroelectric material in 2D form is categorically necessitated. Intriguingly, the atomically thin nature of 2D heterostructures can also pave the way to realize atomic scale modulation of the local electronic properties across the material's edges, as has recently been used to demonstrate sub-nm gate length in vertical MoS$_2$ devices[21].

Among 2D ferroelectrics that are characterized by the presence of exclusively out-of-plane (OOP) or in-plane (IP) polarization such as in CuInP$_2$S$_6$[22], 1T-WTe$_2$[23], SnTe[24] etc., In$_2$Se$_3$ has drawn special attention owing to the existence of stable intercoupled IP and OOP ferroelectricity down to the monolayer limit[25–33]. The origin of the ferroelectricity in α-In$_2$Se$_3$ arises from the relative displacement of the central Se atomic layer with respect to the adjacent In atomic layers, which breaks the centrosymmetry in the crystal and results in two energetically degenerate polarization states[31]. This unique character makes In$_2$Se$_3$ a potential candidate for emerging artificial intelligence, information processing and memory applications. Moreover, its high optical absorption[34], strong photoresponse[35] and phase-dependent visible to infrared bandgap[36] become advantageous from an optoelectronics point-of-view.

Herein, we study the optoelectronic characteristics of vdW heterostructures based on ferroelectric In$_2$Se$_3$ and semiconducting MoS$_2$. In particular, the device channel is made from either monolayer or few layers MoS$_2$, while a thin (~ 50 nm) layer of In$_2$Se$_3$ is inserted beneath half of the device channel in between the MoS$_2$ and the SiO$_2$ dielectric (Fig. 1 a, b).

Our experiments exhibit that selective poling of the In$_2$Se$_3$ can significantly modulate the conduction in the MoS$_2$ channel leading to distinct "On" and "Off" states with a maximum ratio of ~ 10$^4$ and 50 at V$_G$ = - 20 V and 0 V, respectively. In addition, nonvolatile multilevel photoresponse is demonstrated following different poling conditions with the highest photoresponsivity of ~ 1275 A/W following an applied negative gate voltage of - 90 V, which is higher than other reports using either MoS$_2$ or In$_2$Se$_3$ based optoelectronic devices[7,8,29,37–41]. Using Kelvin probe force microscopy (KPFM), we show that the observed modulation in the device conductivity corresponds to a narrow potential modulation across the device channel caused by the α-In$_2$Se$_3$ poled edge (Fig. 1c,d). In particular, the intercoupled ferroelectric polarization actively actuates the channel carrier concentration in the MoS$_2$ leading to non-volatile switching between a diode like *n-i* and an *n-i-n* junctions. The latter is rationalized by a novel edge induced junction mechanism that can potentially enables the realization of FeFETs with ultra-short semiconductor channels, similar to recently reported device architecture[21]. Finally, the heterostructure FeFETs based on monolayer MoS$_2$ exhibit non-volatile poling dependent photoluminescence (PL) of the bright A- and B- excitonic channels that are rationalized by the induced changes in the heterostructure electronic bands alignment. The herein demonstrated dipole modulated FeFETs can be utilized to fabricate functional multimodal optoelectronic memory devices with potentially sub-nm narrow channel dimensions having far-reaching potential for logic, memory, neuromorphic and optoelectronic applications.

## Results –

Few layers of α-In$_2$Se$_3$ were first exfoliated onto degenerately *p*- doped Si substrate capped with 300 nm thermal oxide by a standard scotch tape method[42]. Similarly, few layers of MoS$_2$ were exfoliated onto SiO$_2$/Si and then deterministically transferred using a dry viscoelastic stamp[41] over the α-In$_2$Se$_3$ flake such that only half of the device channel is placed over the In$_2$Se$_3$ (Fig. 1a,b). E-beam lithography was used to fabricate source and drain contacts while the underlying *p*-doped Si was used as back-gate. The atomic configuration of the In$_2$Se$_3$-MoS$_2$ heterostructure following the application of negative and positive back gate potentials are shown in Figure 1a and 1b, respectively. In particular, the atomic arrangement of the In$_2$Se$_3$ layer is modified by the different polarity of the applied field leading to alternating polarization states. For negative OOP poling, the central Se atom is vertically translated towards the top In atom leading to net OOP polarization in the downward (↓) direction as shown by the red arrow. For positive poling, the central Se atom vertically translates towards the bottom In atom leading to net OOP polarization in the upward (↑) as indicated by the red arrow. The IP polarizations are also subsequently modified as a result of the coupled nature leading to varying polarizations (→ and ← respectively). The intercorrelated IP and OOP polarizations are spread out across the entire structure of the flake. Such dipolar variations induce charge carrier migration on the overlying MoS$_2$ leading to the formation of local regions with different charge carrier concentration (insets Fig. 1a, b). For negative poling, the OOP dipole (oval with green body) is directed such that the negative charge center is in proximity to the overlying MoS$_2$, leading to repulsion of electrons and formation of an intrinsic '*i*' type region, which is verified by the measured potential profile shown in Figure 1c. For positive poling, the specific dipolar arrangement with the positive charge center in proximity to the overlying MoS$_2$ attracts electrons leading to the formation of an electrons doped '*n*' type region (Fig. 1d). Intriguingly, in this configuration, the specific orientation of the IP dipoles (oval with red body) leads to the formation of a narrow '*i*' type region, right across the material edge (discussed in detail below). Figure 1e shows the AFM topography of the heterostructure device and its height profile along the red dotted line. The height of the exfoliated MoS$_2$ is ascertained to be ~ 25 nm, whereas the cumulative height of the In$_2$Se$_3$/MoS$_2$ structure is ~ 75 nm (i.e., the thickness of the In$_2$Se$_3$ is 50 nm). Figure 1f shows the Raman spectra obtained from various regions in the heterostructure device and its optical image (inset image). The Raman spectra for the MoS$_2$ region has two distinct peaks at ~ 385 cm$^{-1}$ and 407 cm$^{-1}$, which correspond to the E$^1_{2g}$ and A$_{1g}$ modes[5], respectively. Over the heterostructure region, along with the peaks for MoS$_2$, additional peaks are observed at 105 cm$^{-1}$, 178 cm$^{-1}$ and 192 cm$^{-1}$ that correspond to A$_1$(LO + TO), A$_1$(LO) and A$_1$(TO) phonon modes in α-In$_2$Se$_3$[27,43,44], respectively. The presence of the A$_1$(LO) and A$_1$(TO) peaks caused by the LO − TO splitting indicate the lack of inversion symmetry and the ferroelectric nature in the α-In$_2$Se$_3$[29] crystal. The Raman spectrum of the pre-transferred α-In$_2$Se$_3$ is also shown in the figure for comparison.

Figure 2 presents the cross-sectional high-resolution scanning transmission electron microscopy (STEM) image and the energy-dispersive X-ray spectroscopy (EDS) maps of the heterostructure device. The atomic-scale crystalline structure and the defect free clean interfaces in the heterostructure reveal the sample quality (Fig. 2b,c,f). Evidently, the In$_2$Se$_3$ edge was composed of multiple height steps, that results in an extended junction region. The elemental composition mapping also acquired at the edge part (Fig. 2d,e) and at the layer-on-layer heterojunction region (Fig. 2f) to show the nature of contact. The false-colour coded elemental distributions were created by overlapping each individual one on the HADAAF image, which distinctly exhibiting the active part of device consists of a top few-layer MoS$_2$ which is partially covering the underneath In$_2$Se$_3$.

The ferroelectric polarization in 2D In$_2$Se$_3$ and its subsequent control by gate voltage can be utilized to modulate the electronic band structure of the FeFET[45] and consequently the optoelectronic characteristics of the adjacent MoS$_2$ semiconducting channel. To investigate such dipole induced non-volatile device characteristics, electrostatic poling of the In$_2$Se$_3$ was performed by applying a particular gate bias voltage pulse for a duration of 30 sec followed by the withdrawal of the gate voltage and assessment of the current voltage characteristics of the heterojunction device. Figure 3a presents the measured current-voltage output characteristics of the FeFET following three different poling conditions, viz., unpoled and poled at ± 90 V (measurements took place 2 min after the withdrawal of gate bias voltage to allow for discharge of induced electronic trapped charge and stabilization of the current level as will be discussed further below). Such relatively high switching voltages required for complete polarization switching are attributed to the 300 nm SiO$_2$ dielectrics that results in weaker induced vertical electric field across the device channel[29,45]. A significant modulation of the channel conductance is clearly observed due to the different poling conditions. In particular, following positive (negative) gate poling, in which the polarization in the underlying α-In$_2$Se$_3$ points upwards (downwards) the MoS$_2$, the conductivity of the channel decreases (increases) significantly, realizing two distinct stable conduction states i.e., "On" and "Off" states. In addition, the I-V profile shows a diode like behavior following negative gate poling, which is attributed to the junction formation between the $In_2Se_3$ supported- and the unsupported $MoS_2$ regions, as will be discussed below. More intriguingly, a counterintuitive low channel conductance state following positive gate poling is observed, whereas an increase in electron concentration in the $MoS_2$ channel was expected considering the obtained dipole polarization. Such device characteristics are rationalized by a novel edge state actuation mechanism leading to an abrupt potential barrier at the center of the device channel, as will be further discussed below.

Figure 3b shows the transfer characteristics of the device for different gate voltage sweep ranges for $V_D$ = 2 V. A progressively increasing clockwise hysteretic behavior (see supplementary section) is observed with respect to the applied sweep range (inset of Fig. 3b). Such transfer characteristics are attributed to the ferroelectric nature of α-$In_2Se_3$[15,46], and can be used to substantially affect the conductivity of the $MoS_2$ channel ( up to ~ 4 orders of current magnitude for an applied gate potential of - 20 V), thereby realizing non-volatile memory operation (note that the measured current values are significantly larger in the "On" state i.e., ~ 10 μA, in comparison with typical current levels for $In_2Se_3$ based FET devices having similar geometry[29,47–49], which are in the nA - pA range, indicative for the negligible current flow through the $In_2Se_3$ layer).

To evaluate the retention and stability of the "On" and "Off" states in the FeFETs, time resolved current measurements following alternating poling conditions were monitored. The retention characteristics were measured following 30 sec gate voltage pulses of ± 90 V and are depicted in Figure 3c. Additional figures with linear time scales are included in the supplementary section (Fig. S1a, b). As evident from the plots, it is clear that the current stabilizes and follows a linear trend while maintaining two distinct conduction states for a read voltage of 0.1 V with negligible $I_{on}/I_{off}$ degradation for at least one hour. By linear extrapolation of the stabilized current levels in the linear time scale[50–53], it can be argued that the ratio of the two conduction states is expected to remain ~ 10, even after a time period of more than a year, showing great potential for long term data retention. Additionally, device endurance was characterized by recording the current value for an applied drain bias voltage of 0.1 V, 2 min following the application of alternating gate poling voltages of ± 90 V for over 100 cycles (Fig. 3d). The current level for both $I_{on}$ and $I_{off}$ states exhibit no sign of deviation, retaining an on/off ratio of ~ 60 following 100 cycles of alternating poling conditions. Our results show that by controlling the polarization state of the ferroelectric $In_2Se_3$, it is possible to use the FeFET as a non-volatile memory switch. It is to be acknowledged that the hysteretic behavior of the operational FeFET device may stem from contributions of interfacial charge traps as well as ferroelectric switching as evident by the current variations during the first ~ 20 minutes following gate voltage withdrawal (Fig. 3c). Nevertheless, while the quantification of individual contributions from traps and ferroelectric polarization is beyond the scope of this paper, the retention studies indicating stable, non-volatile conductive states after both negative and positive poling conditions that are well distinguished from the current variations during the first 20 minutes (where traps seem to play their major role). Similarly, the conformity of the memory window following hundred cycles of continuous positive and negative poling suggest strongly that the ferroelectric switching effect dominates over charge trapping in our device[54,55].

The observed diode-like characteristics following negative gate voltage actuation suggests the formation of a lateral *n-i* junction across the $MoS_2$ channel[56], which makes it a potential candidate for photodetection applications. To assess the functionality of the FeFET for photodetection applications, polarization dependent photoresponse of the FeFET were studied where the device was subjected to an alternating illumination using a white light source (420 - 720 nm; methods section) following different gate pulses at a drain voltage of 2 V (Fig. 4a). Figure 4b shows a zoomed in section (marked by the black dotted box in Fig. 4a) from the current vs. time plot. It is evident that the calculated photocurrent (see supplementary Fig. S2) is strongly affected by the induced polarization. In addition, while a relatively modest photoresponsivity of ~ 1.46 A/W is achieved following a positive gate pulse of + 90 V, a maximum of ~ 1275 A/W is recorded following negative gate pulse of - 90 V (Fig. 4c). Such extraordinary increase (~ 3 orders of magnitude) in photoresponse is attributed to the formation of a lateral *n-i* junction across the device, as will be further discussed below. A comparison of the photoresponsivity in $MoS_2$ based devices having various geometries have been tabulated in Table T1 in the supplementary section. The calculated detectivity similarly presents relatively high values of ~ $12 \times 10^{11}$ Jones following a gate pulse of - 90 V. The photo switching for each individual poling condition and the corresponding rise and fall times are presented in supplementary Figure S3. We note that while current variations following positive gate poling are observed during the beginning of the photo switching measurements similar to the retention analysis and presumably due to de-trapping effect, it has a negligible effect on the relative photocurrent. This suggests that charged traps are mainly introduced below the In$_2$Se$_3$ film and away from the heterojunction interface.

In order to have a quantitative understanding of the underlying conduction modulation mechanism, local surface potential measurements by frequency modulated KPFM were conducted. Figures 5a and 5b show the surface potential and topography profiles across the MoS$_2$ channel taken 2 mins following two extreme poling conditions of - 90 V and + 90 V, respectively for 4 consecutive cycles (the gate, source and drain contacts are grounded throughout the scan). The source and drain electrodes are marked with a golden yellow box. The non-zero potential measured above the drain contact (above the heterostructure) stems from an incomplete potential screening of the induced dipoles below the metal electrode[29,57]. Additionally, the slight potential variations between the different consecutive cycles are attributed to a small drift in the scanning location with respect to the device channel. Figures 5c and 5d show the surface potential maps of the FeFET following withdrawal of - 90 V and + 90 V applied gate voltages showing distinct *n* and *i* regions. It can be seen that the surface potential across the un-supported MoS$_2$ section remains almost constant following both positive and negative gate poling, whereas a clear change in surface potential is observed atop the heterostructure due to different induced polarity. The measured potential maps demonstrate the stability of the induced variations throughout the scan indicating that the main effect can be attributed to the ferroelectric polarization as opposed to charge trapping. Intriguingly, an abrupt potential modulation was observed at the center region of the device channel right at the onset of the heterostructure section following positive gate poling. Such potential variations correspond to the formation of a lateral *n-i-n* junction as can been seen in Figure 5d. A possible explanation for such junction formation is schematically described in Figure 5f, where due to the positive poling, the OOP dipoles in the α-In$_2$Se$_3$ orient such that the polarization is directed towards the hetero-interface, as denoted by the red arrows in Figure 5f. This orientation of the dipolar field corroborates to the accumulation of negative charges in the MoS$_2$. Additionally, the intercoupled IP polarization results in an abrupt electron depletion right across the In$_2$Se$_3$ edge leading to a local potential barrier at the center of the MoS$_2$ channel. Such potential modulation results in a robust "Off" state as seen by the current voltage profile in Figure 3. In contrast, following a negative gate poling, when the dipoles orient such that, they are directed away from the α-In$_2$Se$_3$-MoS$_2$ hetero-interface (denoted by red arrows in Fig. 5e), an energy gap of 1.95 eV was measured across the junction. Such dipolar orientation depletes negative charges in the MoS$_2$ leading to the formation of an *n-i* junction (Fig. 5e), in agreement with the diode-like characteristics shown in Figure 3a and the excellent photoresponsivity that is observed for such poling conditions. The relatively wide potential barrier (FHWM of 1.5 µm) following positive gate actuation (Fig. 5b) is attributed to the multiple $In_2Se_3$ edge steps seen by the cross-sectional HRTEM analysis as well as to the multi-layer $MoS_2$ (~ 25 nm thick) film that introduces electrostatic screening in the vicinity of the junction. This implies that a structure comprising monolayers of $In_2Se_3$ and $MoS_2$ may result in a significantly sharper potential variation, which is of great promise for nanoscale actuation.

The presence of alternating dipolar field at the interface between $In_2Se_3$ and $MoS_2$ can potentially impact the PL of monolayer $MoS_2$. Figure 6a shows the PL spectra measured atop the heterostructure section fabricated with monolayer CVD $MoS_2$ at various applied back-gate voltages and Figure 6b shows the contour map of the PL intensities after normalization with respect to the A- exciton peak. Interestingly, a significant enhancement of the PL signal is observed for negative poling that is characterized by a relatively lower electron concentration in the $MoS_2$ channel. Similar PL characteristics are also obtained following 10 minutes after gate voltage withdrawal indicating the potential for efficient PL emission switching and memorization (Fig. 6c and Fig. S4a). Similar PL variations were observed in $MoS_2$ based field effect devices[58] and in ferroelectric Lithium Niobate ($LiNbO_3$) based $MoSe_2$ heterostructures[59]. In addition, a slight redshift of the A- exciton is observed at positive applied back gate potential and is attributed to the increase in negatively charged Trion concentration[60–62]. Finally, a significant enhancement of the intensity ratio of B- to A- excitonic emission ($I_B/I_A$) is observed for negative poling (see supplementary Fig. S4b). The observed PL modulation of the A- and B- excitonic emission can be explained based on the different electronic band alignment of the $MoS_2$-$In_2Se_3$ heterostructure for upward and downward induced polarizations[63] (Fig. 6d). In particular, the heterostructure experiences a clear transition from staggered to straddling band alignment facilitating clear PL bleaching of the photo-generated electron-hole pairs following positive applied gate voltage (Fig. 6c right). In contrast, following negative gate voltage the band alignment favors continuous injection of photogenerated holes into the $MoS_2$ monolayer facilitating enhanced PL emission of both A- and B- excitons (Fig. 6c left). Thus, the increase $I_B/I_A$ ratio following negative gate poling can potentially be explained by the higher induced hole concentration enabling reduced lifetimes of the excited excitonic states, which supposedly has higher impact on the B- excitonic emission channel due to its lower intrinsic quantum yield compared to the A exciton state[65]. Another plausible explanation could be due to the intimate coupling between the MoS$_2$ layer and the induced dipole field in the In$_2$Se$_3$ that is not sufficiently screened due to the low carrier concentration in the MoS$_2$ following negative gate poling, in comparison with stronger dipole screening following positive gate poling[64–66]. Nevertheless, additional experimental and theoretical work are required to better understand the fundamental nature of the PL modulation of ferroelectric supported MoS$_2$ monolayer system.

## Conclusions –

In summary, non-volatile control over the optoelectronic properties in α-In$_2$Se$_3$-MoS$_2$ based vdW FeFETs is demonstrated. The gate controlled ferroelectric polarization can efficiently influence the carrier concentration in the overlying MoS$_2$, leading to the formation of *n-i-n* and *n-i* junctions following positive and negative poling, respectively. Intriguingly, an abrupt potential modulation across the heterostructure edge can be induced by the IP dipole polarization of the In$_2$Se$_3$, making it potentially suitable for sub-nm semiconductor channeled FeFET comprising non-volatile memory characteristics. The local charge carrier engineering and subsequent junction formation has been verified by surface potential measurements based on FM-KPFM. In addition, superior photoresponse of ~ 1275 A/W under white light illumination has been achieved owing to the efficient separation of photogenerated carriers by the formation of an *n-i* junction. Finally, distinct non-volatile modulation of the bright A- and B- excitonic emission channels has been demonstrated.

## Methods -

**Sample preparation -**

Mechanical exfoliation of α-In$_2$Se$_3$ (2D Semiconductors) and MoS$_2$ (Manchester Nanomaterials) was conducted with crystals of ~ 4-6 mm diameter. For both materials, the first few layers of the bulk crystal were mechanically cleaved to discard native oxide layers on the surface and then the pristine bulk was consequently cleaved to exfoliate flakes. The exfoliated flakes were then transferred onto a pre-patterned degenerately *p*-doped silicon substrate with 300 nm thermal Si oxide. A two-stage dry transfer method was used to fabricate the heterostructure using a viscoelastic Polydimethylsiloxane (PDMS) – Polypropylenecarbonate (PPC) stamp fitted on a three-axis micrometer. To pick up the desired flake, the stamp was brought in contact with the flake and was heated at 50 °C to facilitate adhesion. Next, the stamp with the MoS$_2$ facing downwards was brought above the substrate with the pre-exfoliated In$_2$Se$_3$ and was then slowly brought in contact with the substrate. The substrate was then heated to 100 °C to allow adhesion between the 2D materials and the stamp was slowly lifted after cooling down back to room temperature leaving the MoS$_2$ in contact with the underlying In$_2$Se$_3$.

**Growth and wet transfer of CVD grown MoS$_2$-**

MoS$_2$ monolayers were grown by a space confined CVD approach[67,68]. In a typical growth process, MoO$_3$ (99.5%, Sigma Aldrich) and sulfur powders (99.95%, Sigma Aldrich) were used as the metal and the chalcogen precursors, respectively. A ceramic boat containing ~ 3.5 mg MoO$_3$ powder was placed at the center of a CVD furnace of 1-inch diameter. Following, a Si substrate with 300 nm thermal oxide was mounted on the same boat with its polished surface facing upwards. Few small pieces of mica sheets were also placed above the target substrate. The sulfur (~ 350 mg) boat was placed ~ 22 cm upstream from the MoO$_3$ source-growth substrate. 250 sccm of highly pure Ar (5N) carrier gas was purged into the quartz tube for 10 mins to initiate the process. The furnace temperature was then ramped up to 750 °C at 15 °C per minute rate with a flow of 30 sccm of Ar while the chalcogen was separately heated to 180 °C. The growth time was kept for 5 - 10 mins at 750 °C. MoS$_2$ samples were then transferred onto the desired substrate by a wet transfer process which involved spin coating the MoS$_2$ monolayers with 0.5 % wt polystyrene (PS) [450 mg of PS (mol. wt 280,000 g/mol) in 5 ml of Toluene] for 60 seconds at 3000 rpm.

Following, the substrate was subjected to a two-step baking process: at 90 °C for 30 mins and 120 °C for 15 mins. In order to delaminate the MoS$_2$/PS assembly from the substrate, a surface energy assisted transfer technique was adopted, where a drop of water was poured on one of the exposed edges of the substrate, instantly releasing the assembly. Thereafter, the MoS$_2$/PS film was fished out with the Si/SiO$_2$ substrate and baked with the same previous conditions. Finally, toluene was used to dissolve the PS film.

**Cross-sectional HRTEM and EDS mapping –**

The fabricated heterostructure device was investigated by high-resolution transmission electron microscopy (HRTEM). The TEM imaging and elemental distribution mapping were carried out by utilizing a double Cs-corrected HR-S/TEM, Titan Themis G2 60-300 (FEI/Thermo Fisher, USA), equipped with a Dual-X detector (Bruker Corporation, USA) EDS probe. The EDS elemental spectra from selected region were recorded, processed, and analyzed by Velox software (Thermo-Fisher, USA). The cross-section of the specimen lamella for TEM analysis was prepared by using standard focused ion-beam (FIB) lithography approach. The dual mode, Plasma Focused Ion Beam (PFIB) tool (Helios 5, Thermo Fisher Scientific) equipped with an inductively coupled plasma (ICP) and gas injection system (GIS) was used for site specific, in-situ lamella preparation. First, a layer of tungsten (W) was deposited to protect the sample from possible damage during FIB milling during lamella preparation. Then, the area of interest (lamella) was cut-off by FIB fixed at 54° incident angle and lift-out by utilizing a piezo driven W-needle to transferred on a clamp of special TEM grid, where it successfully welded by standard *in-situ* beam induced deposition technique. Afterwards the final thinning of lamella was carried out by FIB polishing from both sides under different incident angles.

**Device fabrication –**

Standard e-beam lithography [Raith-eLine] followed by electron beam evaporation [Evatec BAK 501A] of 5 nm of Cr and 50 nm of Au were used to fabricate the contact electrodes. Prior to the metal deposition, the sample was subjected to a mild oxygen plasma to remove unwanted resist residues [Low Pressure Plasma System – Diener PCCE] for ~ 5 seconds. The metal deposition rate was set to ~ 0.5 Å/s for Cr and ~ 1 Å/s for Au at the base pressure of ~ 7 x 10$^{-7}$ torr.

**Surface and optoelectronic characterization -**

Atomic force microscopy (AFM) and Kelvin probe force microscopy (KPFM) measurements were conducted in an $N_2$ filled glovebox ($H_2O$ and $O_2$ content < 1 ppm) (Dimension-Scanassist, Bruker Inc.) by frequency modulation (FM-KPFM) technique using conductive Pt/Ir-coated cantilever [PPP-EFM-50, NANOSENSORS$^{TM}$ with ~ 25 nm tip radius]. Semiconductor parameter analyzer [Keysight B1500A] and a probe station equipped with an optical microscope were used to electrically characterize the heterostructure FETs at room temperature in ambient atmosphere. A white light source (spectral range 420 nm - 720 nm) with a uniform intensity of ~ 332 µW/cm$^2$ was used to characterize the effect of multilevel photo response due to induced dipole polarizations. The light was homogeneously illuminated onto the whole device regime.

**Spectroscopic characterization –**

Raman and PL spectroscopy were used to characterize the individual 2D materials as well as their heterostructures using WITec Alpha 300R Raman Microscope in confocal mode comprising of 532 nm laser. A 100x objective (NA = 0.9; Dl ~ 360 nm, 600 g*mm$^{-1}$ grating) was used to focus the laser beam, while keeping the excitation power at ~ 1 mW to avoid degradation of the material.

## Associated Content -

## Supplementary Information -

## Author Information -

**Corresponding author –**

Elad Koren - *Nanoscale Electronic Materials and Devices Laboratory, Faculty of Materials Science and Engineering, Technion - Israel Institute of Technology, Haifa, 3200003, Israel.*

orcid.org/0000-0001-7437-7124; Email – eladk@technion.ac.il

**Authors -**

Debopriya Dutta - *Nanoscale Electronic Materials and Devices Laboratory, Faculty of Materials Science and Engineering, Technion - Israel Institute of Technology, Haifa, 3200003, Israel.*

orcid.org/0000-0003-2086-8336

Subhrajit Mukherjee - *Nanoscale Electronic Materials and Devices Laboratory, Faculty of Materials Science and Engineering, Technion - Israel Institute of Technology, Haifa, 3200003, Israel.*

orcid.org/0000-0003-2674-1264

Michael Uzhansky - *Nanoscale Electronic Materials and Devices Laboratory, Faculty of Materials Science and Engineering, Technion - Israel Institute of Technology, Haifa, 3200003, Israel.*

Pranab Kishore Mahapatra - *Department of Materials Science and Engineering, Tel Aviv University, Ramat Aviv, Tel Aviv, 6997801, Israel.*

Ariel Ismach - *Department of Materials Science and Engineering, Tel Aviv University, Ramat Aviv, Tel Aviv, 6997801, Israel.*

orcid.org/0000-0002-4328-9591

**Author Contributions –**

D.D. performed the experimental work, S.M. and M.U. provided experimental support, P. K. M. performed CVD growth of 2D samples. A.I. and E.K. supervised the work. All authors participated in manuscript writing.

**Conflict of Interest -**

The authors declare no conflict of interest.

## Acknowledgements -

D.D. gratefully acknowledges the support of The Miriam and Aaron Gutwirth Memorial Fellowship and the KLA Excellence Fellowship. E.K. gratefully acknowledges the Israel Science Foundation (ISF) grant 1567/18 and the Israel Innovation authority (Kamin) for financial assistance and the RBNI for the nanofabrication facilities. P. K. M. and A. I. acknowledge the generous support from the Israel Science Foundation (ISF), grants 2171/17 and 2596/21, respectively. Authors thank the technical support by the MIKA staff scientists Dr. Yaron Kauffmann and Dr. Galit Atiya for cross-sectional electron microscopy imaging and FIB for sample preparation, respectively. The work made use of the Micro Nano Fabrication Unit at the Technion. We thank Dr. Sivan Refaely-Abramson for fruitful discussions.

# References -

**Figures –**

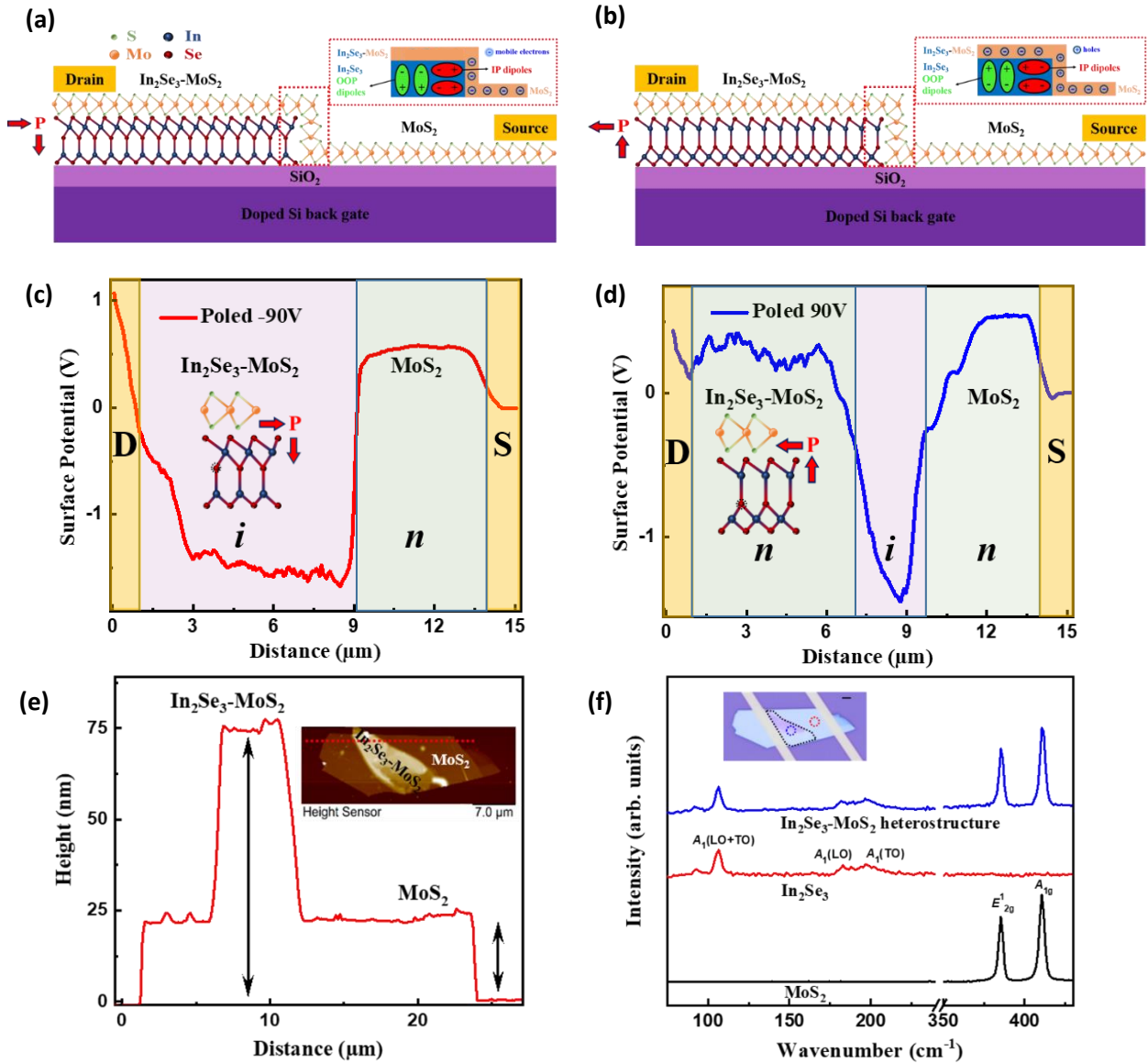

*Figure 1 – Device schematic with dipole induced charge concentration modulation and heterostructure characterization* – Heterostructure schematic with MoS$_2$ transferred onto In$_2$Se$_3$ with corresponding atomic arrangement for **(a)** negative and **(b)** positive poling, respectively. The red dotted box shows the zoomed in charge concentration modulation and migration of electrons in MoS$_2$ as a result of OOP (green oval) and IP (red oval) dipole orientation in In$_2$Se$_3$. The OOP dipoles creates local '*i*' and '*n*' regions in overlying MoS$_2$, depending on the polarization direction indicated by red arrows. Kelvin probe profiles showing local '*p*' and '*n*' regions as a function of **(c)** negative and **(d)** positive poling, respectively. **(e)** Height profile of the device under consideration; inset shows the AFM topography of the device where the red line marks the presented cross-section topography profile, **(f)** Raman spectra of MoS$_2$ flake (black) taken at the position marked by black circular dot, heterostructure (blue) taken at the position marked by blue circular dot and α-In$_2$Se$_3$ (red) taken before the MoS$_2$ transfer (inset showing the optical image of the device). The underlying α-In$_2$Se$_3$ is marked by the dotted green box.

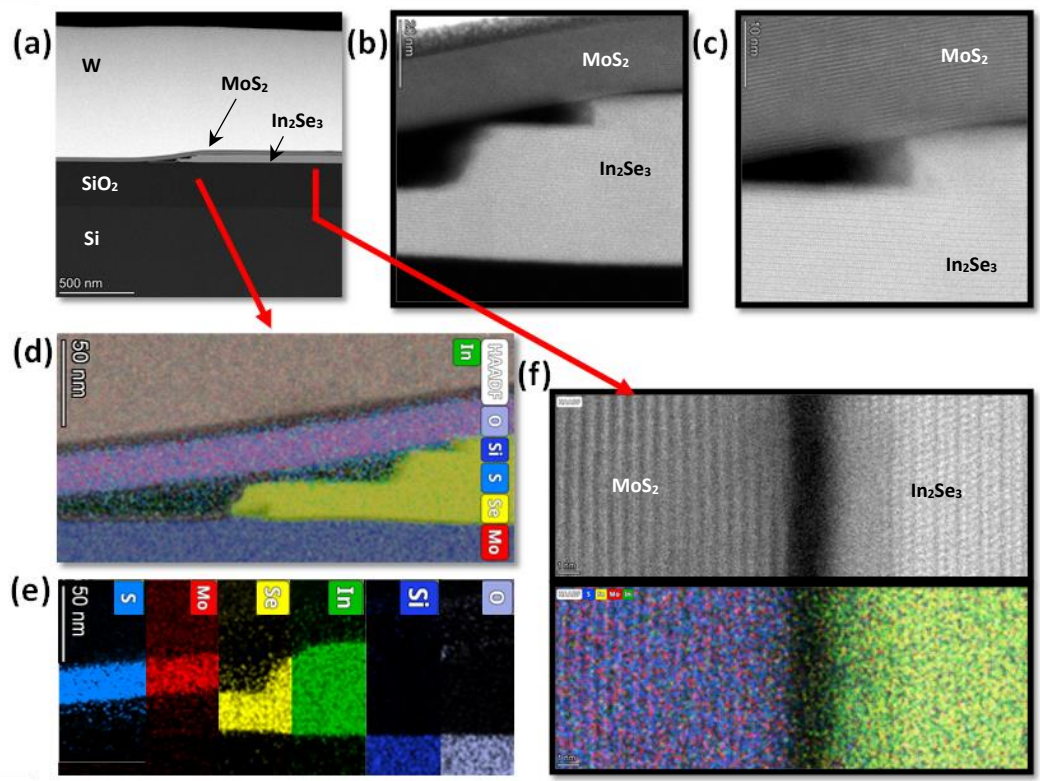

Figure 2 – *Cross-sectional STEM image of the vertically stacked MoS$_2$/In$_2$Se$_3$ layers on a SiO$_2$/Si substrate* – **(a)** Image shows the partially overlapped In$_2$Se$_3$-MoS$_2$ heterostructure. Top bright layer presents the W-metal protection layer, and bottom two dark sections are 300 nm SiO$_2$ and Si, respectively. **(b)** Magnified image at the edge exhibited the nature of contact. **(c)** HRTEM image indicating the crystalline quality and 2D layer structure of both materials. **(d)** STEM/EDS signal mapping of In$_2$Se$_3$-MoS$_2$ heterostructure. **(e)** Individual elemental profiles with false colour codes are combined and depicted for the entire device, where each element is mentioned in the respective inset. Oxygen signal found to be negligible within the MoS$_2$/In$_2$Se$_3$ heterostructure. **(f)** HRTEM image and the corresponding EDS map of the interface of the vertical heterostructure (taken from the marked region), exhibit crystallinity, absence of interfacial defects and elemental distribution without any intermixing, as Mo & S signal only appears for top layer and In & Se only in bottom layer.

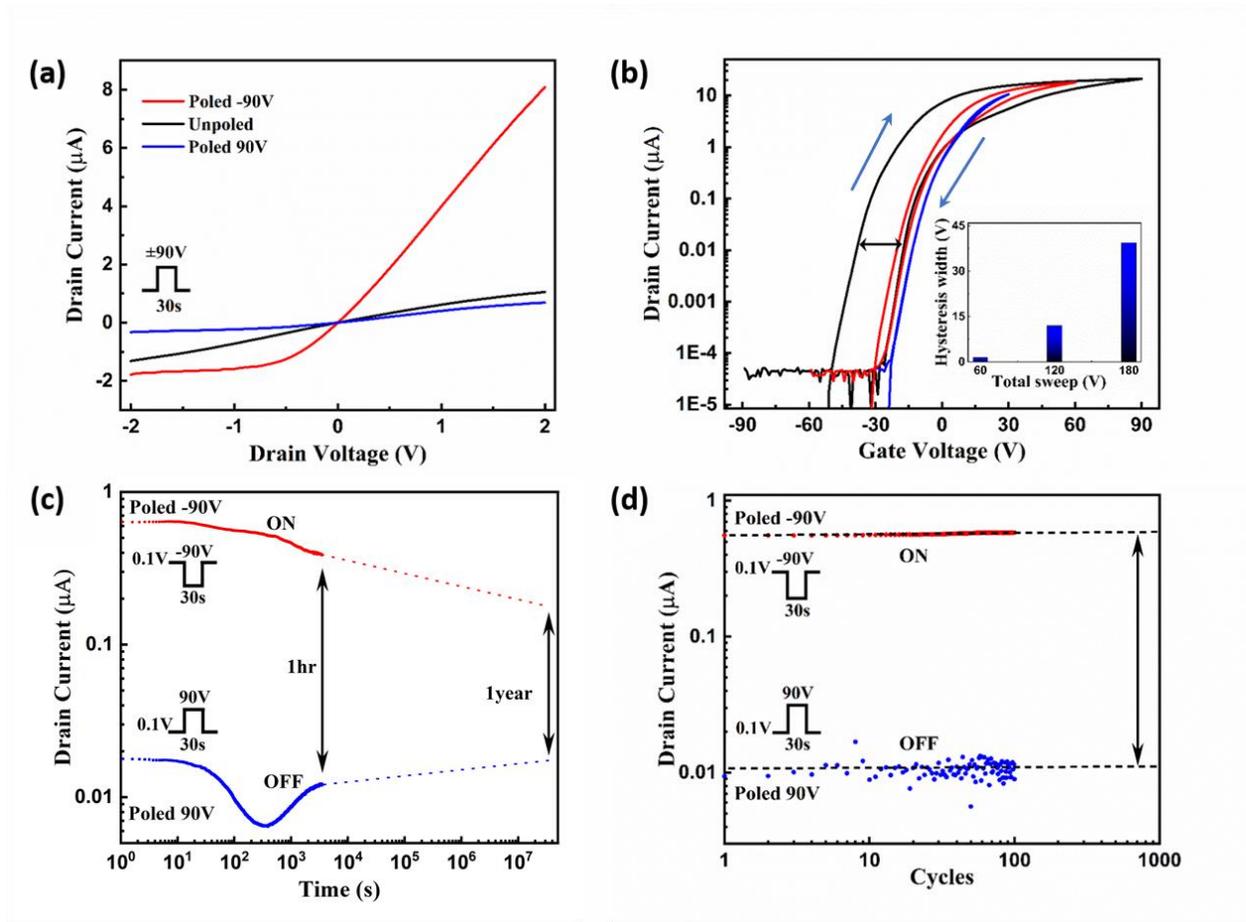

*Figure 3 – Electrical characterization –* **(a)** $I_D$-$V_D$ following different gate poling conditions (gate pulse is indicated in the inset i.e. gate voltage was applied for a duration of 30 sec). **(b)** $I_D$-$V_G$ for different gate range sweeps showing a clockwise hysteresis behavior; inset shows the increasing hysteresis in the device with greater sweep voltage, **(c)** Retention test of the "off" and "on" states measured at $V_G = 0$ V and $V_D = 0.1$ V after respective poling conditions showing expected retention for more than 1 year (see supplementary section for more details). **(d)** Endurance test of the FeFET showing consistent current levels after poling for both "on" and "off" states for at least 100 cycles (gate pulse prior to endurance test and applied drain voltage during the test are indicated in the inset).

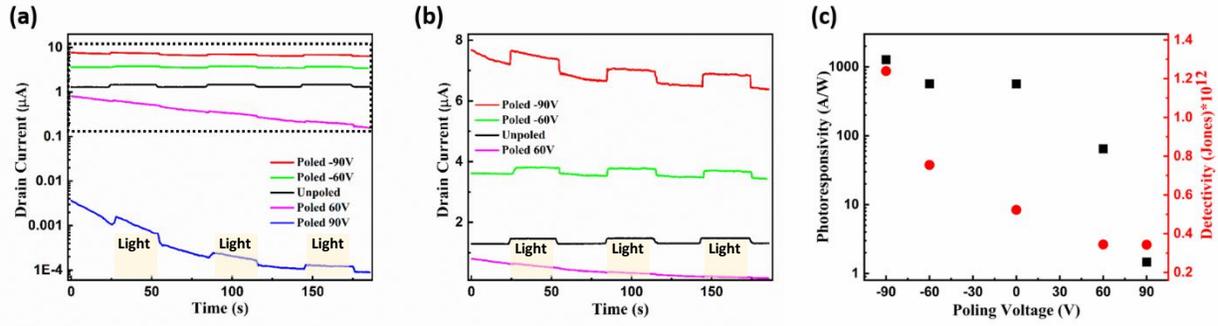

*Figure 4 – Poling dependent photoresponse –* **(a)** Drain current following poling with different gate voltages and under illumination of visible light pulses, **(b)** Drain current for selected poling voltages marked by the dotted black box in (a), **(c)** Calculated photoresponsivity and detectivity for different poling conditions.

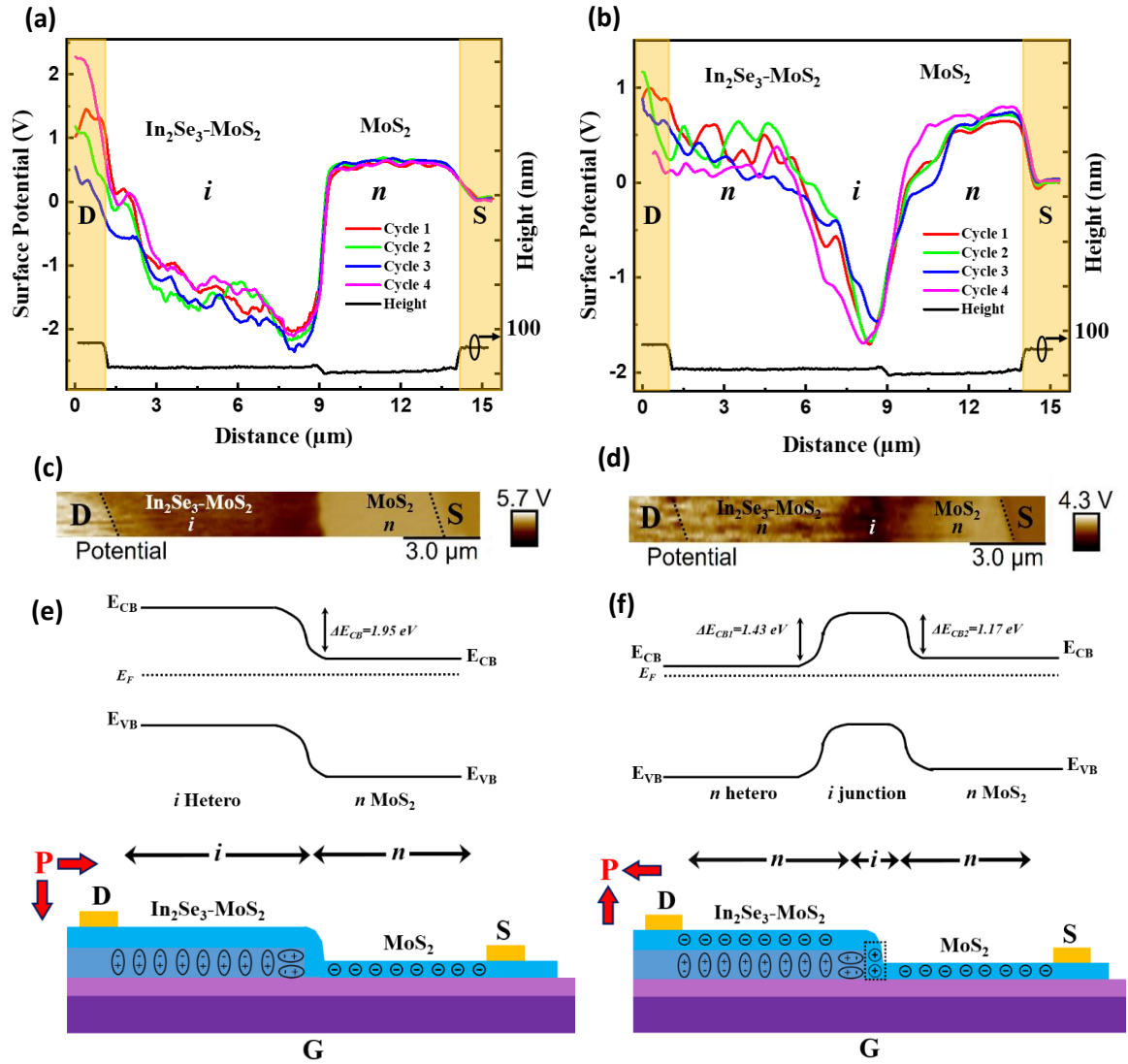

*Figure 5 – Surface potential studies and mechanism of conduction modulation in the FeFET due to the polarization in α-In$_2$Se$_3$* – Potential profiles across the device channel measured 2 mins following multiple withdrawals of applied gate voltages of **(a)** - 90 V and **(b)** + 90 V; the Y- axis on the right shows the height profile. Inset figures show the corresponding crystal orientation of the In$_2$Se$_3$-MoS$_2$ heterostructure part following withdrawal of the respective gate pulses. Surface potential maps of the FeFET following withdrawal of **(c)** - 90 V and **(d)** + 90 V applied gate voltages showing distinct *n* and *i* regions. Schematic illustrations showing the induced dipole polarization within the In$_2$Se$_3$ and the corresponding formation of the *n*, *i* regions in the MoS$_2$ channel according to the extracted band diagrams based on the surface potential mapping following **(e)** negative and **(f)** positive back gate potentials.

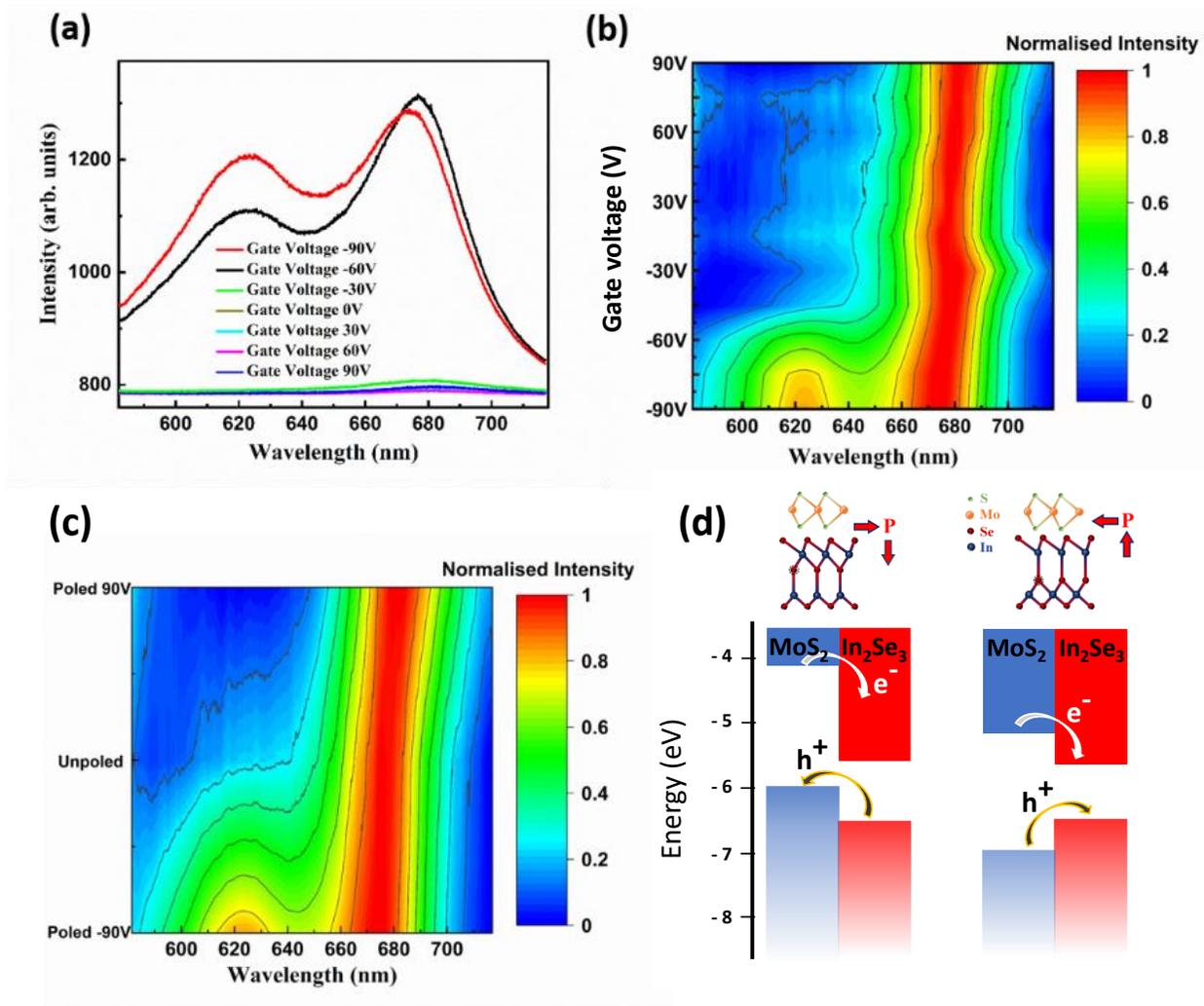

*Figure 6 – Polarization dependent Photoluminescence –* **(a)** PL spectra of α-$In_2Se_3$-$MoS_2$ heterostructure at various gate voltages, **(b)** Contour map of the normalized PL intensities at various applied gate voltages. **(c)** Contour map of the normalized PL intensities, 10 mins following the withdrawal of the applied gate voltage, indicating the retention of the polarization induced PL. **(d)** Schematic description of the electronic band structure alignment in the $In_2Se_3$-$MoS_2$ heterostructure for different induced polarization states following the withdrawal of the applied gate voltage.